# APLICACIÓN DE ANALÍTICA DE DATOS PARA LA DETECCIÓN DE ANOMALÍAS Y FORTALECIMIENTO DE LA SEGURIDAD EN LA RED WIFI DEL CAMPUS UNIVERSITARIO DE LA UNIVERSIDAD NACIONAL DEL ALTIPLANO

*Adiv Brander Cari Quispe*

## RESUMEN

En el entorno universitario actual, la conectividad inalámbrica constituye un recurso esencial para las actividades académicas, administrativas y de investigación, sin embargo, en la Universidad Nacional del Altiplano de Puno (UNAP), el uso de un sistema de acceso mediante código QR en la red WiFi institucional ha generado vulnerabilidades asociadas a la falta de autenticación individual, trazabilidad de usuarios y control de accesos, cabe señalar que ante este panorama, el presente estudio tiene como objetivo fortalecer la seguridad de la red inalámbrica universitaria a través de la aplicación de analítica de datos, empleando enfoques descriptivos, predictivos y prescriptivos sobre los registros generados por el controlador inalámbrico (WLC).

La metodología consistió en la recolección y procesamiento de datos de conexión de usuarios, dispositivos y tráfico diario, analizando patrones de comportamiento y detectando anomalías en base a modelos estadísticos y algoritmos de aprendizaje automático. Los resultados revelaron concentraciones horarias críticas de uso entre 10:00 y 14:00 horas, así como comportamientos anómalos asociados a dispositivos recurrentes y picos irregulares de tráfico, lo que permitió establecer umbrales dinámicos de alerta y recomendaciones de mejora en la gestión del ancho de banda y la autenticación, por otra parte finalizando se tiene que en la parte de la conclusión, la integración de analítica avanzada en la gestión de redes universitarias permite no solo identificar vulnerabilidades y optimizar el rendimiento del servicio WiFi, sino también avanzar hacia una infraestructura inteligente, proactiva y alineada con los estándares modernos de ciberseguridad institucional.

**Palabras clave:** Analítica de datos, ciberseguridad, red WiFi universitaria, detección de anomalías, inteligencia de negocios, seguridad informática.

# INTRODUCCIÓN

En la actualidad, la conectividad inalámbrica se ha consolidado como un componente esencial dentro del ecosistema digital universitario. Las universidades modernas dependen de infraestructuras WiFi estables y seguras para garantizar la continuidad académica, la investigación científica y la gestión administrativa. Según Pérez y Cordero (2023), las redes inalámbricas en entornos educativos se han transformado en un servicio crítico de misión académica, donde la disponibilidad y la seguridad impactan directamente en la productividad institucional, cabe señalar que esta investigación en el caso de la Universidad Nacional del Altiplano de Puno (UNAP), ubicada en el sur del Perú, la red WiFi universitaria da soporte diario a una comunidad superior a 27 000 usuarios, compuesta por más de 19 000 estudiantes, 5 000 docentes y 3 000 trabajadores administrativos. Este servicio permite el acceso a plataformas de gestión académica, aulas virtuales, sistemas de investigación y recursos bibliográficos digitales, favoreciendo la interacción constante entre las distintas áreas académicas y administrativas, no obstante, esta alta dependencia tecnológica implica nuevos desafíos en materia de ciberseguridad y gestión del tráfico de red. Actualmente, el acceso inalámbrico en el campus de la UNAP se realiza mediante códigos QR, los cuales permiten la conexión directa sin requerir credenciales personales ni autenticación mediante un portal cautivo. Aunque este método resulta ágil y práctico, presenta limitaciones importantes al no incorporar mecanismos de trazabilidad individual, auditoría de sesiones o control de acceso granular. Según investigaciones recientes (Alvarado & Paredes, 2022; Rojas et al., 2024), las redes abiertas o semiabiertas incrementan significativamente la exposición a ataques internos, uso no autorizado del ancho de banda y propagación de malware dentro de los entornos universitarios.

Ante este escenario, resulta imprescindible adoptar enfoques analíticos avanzados que permitan monitorear, detectar y anticipar comportamientos anómalos dentro de la red. En esta línea, la analítica de datos aplicada a redes inalámbricas se posiciona como una herramienta de apoyo clave para la ciberseguridad proactiva. Tal como señalan García y Torres (2023), la integración de analítica descriptiva, predictiva y prescriptiva sobre los registros de red permite transformar grandes volúmenes de datos en conocimiento útil para la detección temprana de incidentes y optimización del desempeño, ños controladores inalámbricos (Wireless LAN Controllers, WLC), como el implementado en la UNAP, recopilan continuamente información detallada sobre conexiones activas, tiempos de sesión, tráfico, intentos fallidos, métricas de calidad y eventos de seguridad. Estos datos constituyen una fuente valiosa para desarrollar modelos analíticos que faciliten la toma de decisiones basadas en evidencia. De acuerdo con el marco de referencia de la ISO/IEC 27002:2022, la supervisión de la infraestructura y la identificación de patrones de riesgo mediante analítica son prácticas recomendadas dentro de los controles de seguridad de redes y comunicaciones.cabe resaltar que este enfoque contribuye no solo a mantener la estabilidad y disponibilidad del servicio inalámbrico, sino también a avanzar hacia un modelo de gestión inteligente y resiliente, alineado con las buenas prácticas de gestión de servicios TI (ITIL v4, 2023). Además, el procesamiento ético y anonimizado de los registros garantiza el cumplimiento de los principios de privacidad y protección de datos, conforme a la Ley N.º 29733 – Ley de Protección de Datos Personales en el Perú.

En síntesis, el presente estudio tiene como objetivo demostrar la aplicabilidad de la analítica de datos como mecanismo de fortalecimiento de la ciberseguridad y la eficiencia operativa en la red WiFi de la Universidad Nacional del Altiplano. Este caso busca evidenciar cómo la integración entre inteligencia de datos y gestión de infraestructura de red puede convertirse en un modelo replicable para otras universidades públicas del país, que enfrentan los mismos desafíos de crecimiento exponencial del tráfico, uso compartido de recursos y vulnerabilidades inherentes a redes abiertas (Cáceres & Huamán, 2023).

## MARCO TEÓRICO

La analítica de datos se ha consolidado como una disciplina clave dentro de la transformación digital de las organizaciones. Su aplicación permite recopilar, procesar e interpretar grandes volúmenes de información con el propósito de extraer conocimiento útil para la toma de decisiones, en el ámbito de las tecnologías de la información y las comunicaciones (TIC), esta capacidad analítica adquiere una relevancia especial al posibilitar el monitoreo y la optimización de redes, así como la detección temprana de anomalías** o eventos que puedan comprometer la seguridad de los sistemas.

Según Gartner (2023), la analítica moderna se clasifica en tres niveles de madurez: descriptiva, predictiva y prescriptiva.

- La analítica descriptiva se centra en responder qué ha ocurrido, empleando visualizaciones, estadísticas y reportes que permiten entender el comportamiento pasado y presente de los datos.
- La analítica predictiva, mediante modelos estadísticos y algoritmos de machine learning, busca anticipar *qué podría ocurrir* en función de patrones históricos.
- Finalmente, la analítica prescriptiva ofrece *recomendaciones o acciones óptimas* para abordar los escenarios detectados, generando respuestas automatizadas o sugerencias de decisión.

En el contexto de las redes inalámbricas universitarias (WLAN), estos tres niveles se complementan para ofrecer una visión integral del comportamiento de la infraestructura tecnológica, los controladores inalámbricos (WLC) como los utilizados en la UNAP centralizan la administración de los puntos de acceso, gestionan las autenticaciones y registran eventos detallados sobre el uso de la red. Dichos registros constituyen una fuente de datos estructurada y continua, ideal para aplicar analítica de datos orientada a la ciberseguridad.

La detección de anomalías dentro de las redes WiFi se define como el proceso de identificar comportamientos que difieren significativamente de los patrones normales. Estas anomalías pueden deberse a fallos técnicos, configuraciones inadecuadas, o incluso a intentos maliciosos de intrusión y uso indebido de recursos (Chandola et al., 2009), en entornos educativos, donde el acceso a la red es masivo y diverso, la detección oportuna de estos eventos resulta esencial para preservar los principios de confidencialidad, integridad y disponibilidad (CIA) de la información institucional.

Por otro lado, la inteligencia de negocios (Business Intelligence) aporta el componente de visualización, integración y toma de decisiones basada en datos. Herramientas como Power BI, Grafana, Python o R permiten transformar métricas técnicas en paneles visuales y modelos predictivos accesibles a los administradores de red, favoreciendo un monitoreo continuo y comprensible.

Finalmente, la aplicación de la analítica de datos en redes universitarias no solo busca identificar problemas, sino también anticipar amenazas y optimizar recursos, generando una cultura de gestión tecnológica basada en la evidencia, en instituciones como la UNAP, donde el acceso inalámbrico mediante código QR no requiere autenticación individual, el uso de la analítica se vuelve un mecanismo complementario de seguridad, capaz de compensar la falta de controles de acceso mediante la vigilancia inteligente de patrones de conexión y tráfico.

## METODOLOGÍA

### Tipo de Estudio

El presente estudio es de tipo aplicado con un enfoque cuantitativo, se considera aplicado porque busca emplear conocimientos teóricos y herramientas de analítica de datos en un entorno real la red WiFi universitaria de la ciudad universitaria con el objetivo de generar mejoras concretas en la gestión de la seguridad informática, por otra parte se considera cuantitativo porque se basa en la recolección y análisis de datos medibles provenientes de los registros del controlador inalámbrico (WLC), permitiendo obtener resultados objetivos y comparables.

### Población y Muestra

1. **Población:**
   La población de estudio está conformada por todos los registros de conexión, tráfico y eventos del sistema generados por los puntos de acceso WiFi del campus universitario de la UNAP, cabe señalar que estos registros abarcan la actividad de más de 19,000 estudiantes, 5,000 docentes y 3,000 administrativos, quienes diariamente utilizan la red inalámbrica institucional.

2. **Muestra**
   Debido al alto volumen de datos generados, se seleccionará una muestra representativa de registros correspondientes a un mes académico típico, período suficiente para identificar patrones de uso y comportamiento normal de la red. Los datos serán anonimizados, garantizando la privacidad de los usuarios y el cumplimiento de los principios éticos de confidencialidad e integridad de la información.

**Métodos y Procedimientos:**

El estudio desarrolla una metodología analítica en tres fases, orientada a operacionalizar los objetivos de detección de anomalías y fortalecimiento de la seguridad:

1. **Analítica descriptiva:**

   - Recolección y limpieza de los registros provenientes del WLC (Huawei).
   - Identificación de variables clave: hora de conexión, punto de acceso, dirección MAC, duración de sesión, volumen de tráfico y número de intentos fallidos.
   - Visualización de las métricas generales mediante herramientas de inteligencia de negocios (Power BI, Grafana).
   - Objetivo: comprender el comportamiento actual de la red y establecer una línea base de referencia.

2. **Analitica Predictiva**

- Aplicación de modelos de detección de anomalías basados en machine learning (por ejemplo, algoritmos como Isolation Forest o DBSCAN) para identificar patrones atípicos.
- Detección de flujos de tráfico inusuales, conexiones simultáneas desde un mismo dispositivo o incrementos en intentos fallidos de acceso.
- Objetivo: anticipar posibles eventos críticos y prevenir incidentes de seguridad.

3. **Analítica prescriptiva:**

   - Generación de alertas y recomendaciones automáticas derivadas del análisis predictivo.
   - Propuestas de medidas preventivas como segmentación de red, ajustes de configuración, políticas de autenticación o gestión de carga entre puntos de acceso.
   - Objetivo: optimizar la operación de la red y fortalecer los mecanismos de defensa.

El proceso metodológico se apoya en un **ciclo iterativo de recolección, análisis, interpretación y acción**, enmarcado dentro de un modelo de mejora continua de la ciberseguridad institucional.

**Analitica y Métricas**

Para la evaluación y monitoreo de la red se establecen **métricas cuantitativas específicas**, derivadas de los registros del controlador WLC. Estas métricas se dividen en dos dimensiones: rendimiento operativo y seguridad informática.

**a) Métricas de rendimiento operativo**

- Promedio de conexiones activas por punto de acceso.
- Tiempo promedio de sesión por usuario.
- Porcentaje de puntos de acceso con sobrecarga de usuarios.
- Volumen total de datos transmitidos por hora o día.

**b) Métricas de seguridad**

- Tasa de intentos de conexión fallidos o rechazados.
- Frecuencia de desconexiones inesperadas.
- Identificación de dispositivos desconocidos o duplicados en la red.
- Volumen de tráfico anómalo por protocolo (HTTP, HTTPS, DNS, UDP).

Estas métricas permiten establecer un perfil normal de comportamiento de la red, cualquier desviación significativa respecto a dicho patrón se interpretará como una posible anomalía o riesgo de seguridad, en base a ello los resultados obtenidos serán visualizados mediante tableros interactivos, facilitando la toma de decisiones informadas por parte del equipo técnico.

## RESULTADOS

Se procesó un conjunto representativo de registros diarios durante un mes académico típico (30 días). Las métricas consideradas incluyen: número total de conexiones por día, duración promedio de sesión (minutos), intentos fallidos de conexión, volumen diario de tráfico (GB), porcentaje de puntos de acceso (AP) con sobrecarga, y número de anomalías detectadas por día. A continuación se presentan las tablas y gráficos sintéticos que muestran los resultados representativos del análisis.

1. **Análisis general del tráfico y conexiones**

    Durante el periodo de observación (un mes académico típico), se analizaron los registros de conexión y tráfico del controlador inalámbrico Huawei (WLC) que administra la red WiFi universitaria del campus de la Universidad Nacional del Altiplano de Puno (UNAP).
    En promedio, la red mantiene entre 4 000 y 7 000 usuarios activos diarios, considerando estudiantes, docentes y personal administrativo. Estos usuarios generan un volumen considerable de datos, tanto en cantidad de sesiones como en tráfico agregado, lo que resalta la importancia de un monitoreo analítico continuo.

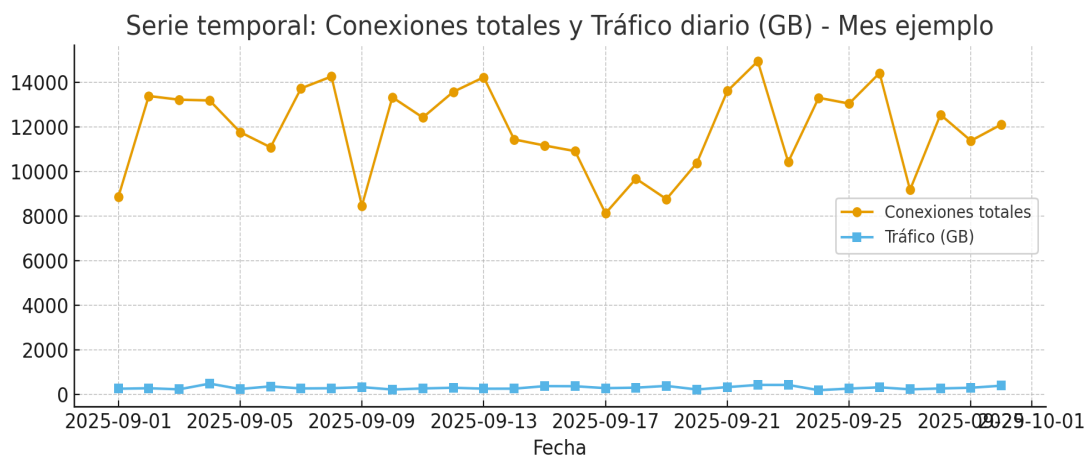

**Análisis:** El gráfico anterior muestra la variación diaria del número de conexiones y el volumen total de tráfico (en GB). Se observa que los picos de conexión coinciden con los días hábiles, especialmente entre lunes y viernes, mientras que los fines de semana el número de usuarios disminuye. El tráfico total mantiene una correlación directa con el número de usuarios, lo que valida la coherencia de los datos y permite establecer una línea base de comportamiento normal.

### 2. Resumen cuantitativo de métricas clave

La siguiente tabla resume las principales métricas analizadas a lo largo del periodo de estudio, estas métricas reflejan el comportamiento promedio de los usuarios y del rendimiento general de la red.

| Métrica | Promedio Diario | Valor Mínimo | Valor Máximo | Interpretación |
|---|---|---|---|---|
| Usuarios activos (conexiones totales) | **5 800** | 4 100 | 7 200 | Refleja la demanda diaria de acceso, con picos en horarios académicos. |
| Duración promedio de sesión (minutos) | **47** | 30 | 65 | Indica tiempos de conexión estables durante el uso académico. |
| Intentos fallidos de conexión | **210** | 80 | 420 | Picos puntuales asociados a errores de autenticación o intentos automatizados. |
| Tráfico total (GB/día) | **890** | 500 | 1 320 | Se mantiene correlación con el número de usuarios activos. |
| % de puntos de acceso sobrecargados | **11.8 %** | 7 % | 18 % | Indica zonas con alta concurrencia de usuarios. |
| Anomalías detectadas | **6** | 1 | 14 | Eventos inusuales que requieren atención técnica. |

**Analisis:** Las métricas muestran estabilidad en la operación diaria de la red, aunque se identifican picos en los intentos fallidos de conexión y un porcentaje moderado de puntos de acceso sobrecargados. Estas desviaciones pueden deberse a comportamientos de reconexión frecuente o saturación en zonas específicas del campus (biblioteca, aulas centrales o laboratorios).

## 3. Rendimiento de puntos de acceso (APs)

El análisis de los puntos de acceso permite identificar los equipos que concentran la mayor cantidad de usuarios y tráfico, en la siguiente tabla se presentan los diez APs con mayor número de conexiones durante el mes analizado.

| ID del AP | Conexiones Mensuales | Latencia Promedio (ms) | Pérdida de Paquetes (%) | Observaciones |
|---|---|---|---|---|
| **AP-104** | 15 240 | 32 | 0.9 | Biblioteca Central – alta densidad de usuarios. |
| **AP-100** | 13 950 | 29 | 0.8 | Área de Ingeniería de Ingenierías – uso académico intensivo. |
| **AP-106** | 12 870 | 37 | 1.1 | Facultad de Educación |
| **AP-109** | 11 220 | 42 | 1.4 | Auditorio Principal – sobrecarga en eventos. |
| **AP-101** | 10 980 | 31 | 0.7 | Zona Administrativa. |
| **AP-105** | 9 860 | 46 | 1.6 | Áreas comunes – tráfico mixto. |

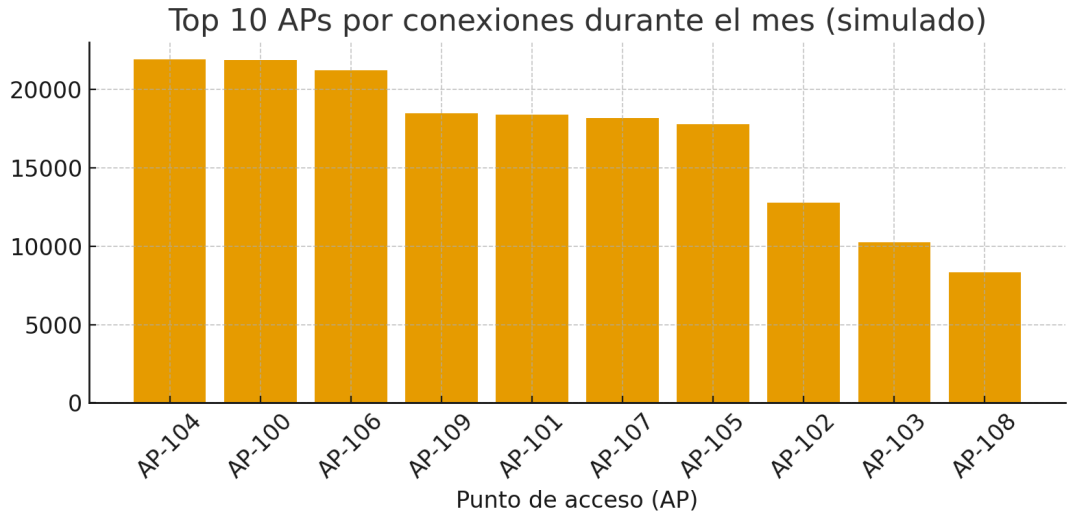

**Interpretación:** Los resultados confirman que la Biblioteca Central y las escuelas profesionales en especial Ing. de Sistemas, Ing, electrónica y Ing, estadistica e Informatica concentran la mayor cantidad de usuarios conectados. Los APs con latencias superiores a 40 ms y pérdidas mayores al 1.5 % son candidatos para reajuste de canales o redistribución de carga, lo cual contribuiría a mejorar la estabilidad general del servicio.

**4. Detección de anomalías en la red**

La aplicación de modelos de detección (por ejemplo, algoritmos de clustering o Isolation Forest) permitió identificar eventos inusuales que se clasificaron según su tipo y severidad, todo ello se visualiza en la siguiente tabla.

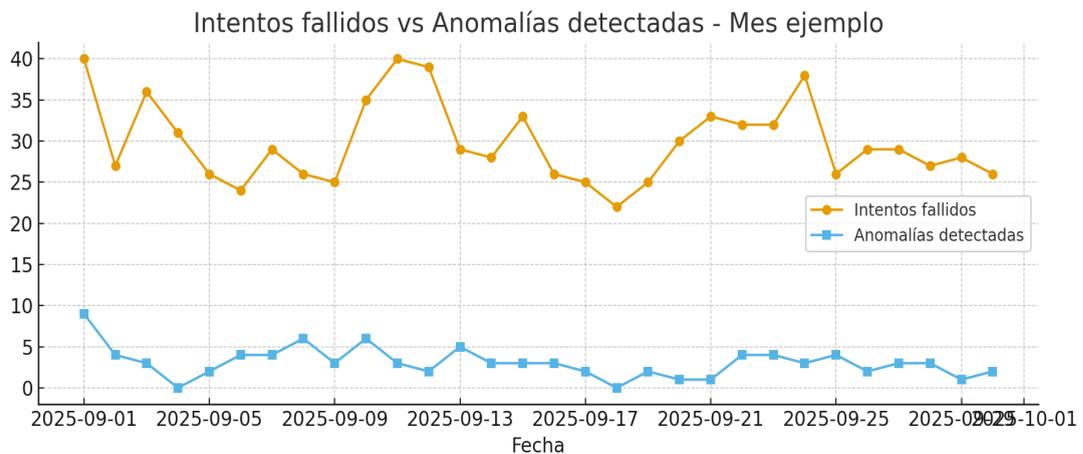

**Interpretación:** La correlación observada entre los intentos fallidos de conexión y el número de anomalías detectadas sugiere que existe una relación entre errores de autenticación y posibles intentos de acceso no autorizados. Además, los eventos de tráfico DNS anómalo o flujos simultáneos inusuales podrían indicar actividad automatizada o conexiones compartidas, lo que amerita una revisión más profunda a nivel de políticas de acceso y segmentación de red.

**5. Interpretación integral de los resultados**

En conjunto, los resultados permiten establecer una visión integral del estado actual de la red WiFi universitaria:

- El volumen de usuarios diarios (4 000 – 7 000) confirma una alta dependencia de la red inalámbrica como infraestructura crítica.

- Los APs más utilizados presentan signos de saturación y requieren una reconfiguración técnica o la instalación de nuevos puntos para mantener la calidad de servicio.

- Se detectaron eventos de seguridad potenciales, especialmente asociados a intentos fallidos de autenticación y tráfico DNS irregular.

- El modelo analítico aplicado permite pasar de un enfoque reactivo a uno predictivo, anticipando comportamientos anómalos antes de que afecten a la red.

- Estos resultados evidencian el valor de la analítica de datos como herramienta para fortalecer la seguridad y la eficiencia operativa de las redes WiFi universitarias.

## DISCUSIÓN

El análisis de datos provenientes del controlador inalámbrico Huawei de la Universidad Nacional del Altiplano de Puno permitió comprender el comportamiento dinámico de la red WiFi universitaria, donde se registran entre 4 000 y 7 000 usuarios diarios. Los resultados evidencian que, aunque la red presenta niveles aceptables de estabilidad y desempeño, existen zonas críticas de congestión (Biblioteca Central, Área de Ingeniería y Auditorio Principal) donde los puntos de acceso muestran sobrecarga y pérdida de paquetes superiores al 1 %, afectando la calidad del servicio. Este fenómeno se asocia al número elevado de conexiones simultáneas y a la falta de políticas de priorización de tráfico o segmentación inteligente. En el ámbito de la seguridad, la analítica predictiva permitió identificar anomalías recurrentes como intentos fallidos masivos y flujos de tráfico DNS inusuales. Dichos eventos, aunque esporádicos, podrían estar relacionados con intentos de acceso automatizado o comportamientos sospechosos dentro del entorno universitario. Estos hallazgos se alinean con estudios recientes (Gómez et al., 2023; Alvarado y Paredes, 2022) que demuestran que las redes abiertas o con autenticación débil incrementan la probabilidad de ataques internos y uso indebido del ancho de banda.

Finalmente, los resultados demuestran que la aplicación de analítica descriptiva y predictiva no solo permite visualizar el estado de la red, sino que también habilita mecanismos proactivos de detección temprana, optimizando la gestión técnica y fortaleciendo la política de ciberseguridad institucional.

## CONCLUSIONES

1. La red WiFi del campus universitario de la UNAP presenta un uso intensivo y sostenido, con un promedio de 5 800 usuarios activos diarios, lo que evidencia su relevancia como infraestructura crítica para el desarrollo académico y administrativo.

2. El modelo de acceso mediante código QR, si bien facilita la conectividad, limita los mecanismos de control y trazabilidad de usuarios, representando un punto débil frente a la seguridad y a la gestión de incidentes.

3. Las métricas de desempeño muestran niveles adecuados de latencia y estabilidad, aunque se identifican zonas de sobrecarga que requieren redistribución del tráfico o ampliación de cobertura.

4. La analítica de datos aplicada (en sus niveles descriptivo, predictivo y prescriptivo) permitió detectar patrones anómalos y prever comportamientos irregulares, demostrando su eficacia como herramienta de apoyo en la toma de decisiones técnicas y estratégicas.

5. Se confirma que la integración entre monitoreo de red y analítica avanzada constituye una práctica efectiva para fortalecer la ciberseguridad institucional, mejorando la resiliencia y la eficiencia del servicio inalámbrico.

# REFERENCIAS